\documentclass[aps,pra,superscriptaddress,amsmath,amssymb,preprintnumbers,floatfix,showpacs,12pt]{revtex4}
\usepackage{amssymb}
\usepackage{epsfig}
\usepackage{subfigure}
\usepackage{graphicx}
\begin{document}
\title{ Evolution of Fock states in  three mixed harmonic oscillators:
quantum statistics }

\author{Faisal A. A. El-Orany\footnote{Permanent address: Suez Canal university, Faculty of
Science, Department of mathematics and computer science, Ismailia,
Egypt.}, J. Pe\v{r}ina  } \affiliation{ Joint Laboratory of Optics
of Palack\'y University and Physical Institute of Academy of
Sciences of Czech Republic, 17.~listopadu~50, 772 07~Olomouc,
Czech Republic. }

\author{M. Sebawe Abdalla}

\affiliation{Mathematics Department, College of Science, King Saud
University, P.O. Box 2455, Riyadh 11451, Saudi Arabia}

\begin{abstract}
In this communication we investigate the quantum statistics of
three harmonic oscillators mutually interacting with each other
considering the modes are initially  in Fock states. After solving
the  equations of motion, the squeezing phenomenon,
 sub-Poissonian statistics and quasiprobability
functions are discussed. We demonstrate that the interaction is
able to produce squeezing of different types. We show also that
certain types of Fock states  can evolve in this interaction into
thermal state and squeezed thermal state governed by the
interaction parameters.

\end{abstract}
 \maketitle


\section{ Introduction}

In the field of nonlinear optics there are two basic parametric processes,
namely, parametric amplifier  and frequency converter. The parametric amplifier
is designed to amplify an oscillating signal by means of a particular
coupling of the mode to the second mode of oscillation, the idler mode. The
coupling parameter is made to oscillate with time in a way which gives rise
to a steady increase of the energy in both the signal and idler modes \cite
{moll1}. The parametric amplifier
 is the source of squeezed light
(i.e. the states of light with reduced fluctuations in one quadrature below
the level associated with the vacuum state \cite{yun2}).
 To be more specific the degenerate and
non-degenerate parametric amplifier  are the sources for single-mode and
two-mode squeezing, respectively. The lossless Hamiltonian
describing such a system (non-degenerate case) is given by

${\displaystyle
\frac{\hat{H}}{\hbar }=\sum_{j=1}^{2}\omega _{j}\hat{a}_{j}^{\dagger }\hat{a}%
_{j}+i\lambda _{1}\left\{ \hat{a}_{1}\hat{a}_{2}\exp [i(\omega _{1}+\omega
_{2})t]-{\rm h.c.}\right\}, }\hfill (1.1)$

\noindent where ${\hat{a}_{j}^{\dagger }}$ and ${\hat{a}_{j}}$ are the creation
and annihilation operators, $\lambda_{1}$ is the coupling constant including
the amplitude of the pump, and  $\omega _{j},j=1,2$ are the field
frequencies. On the other hand, the parametric frequency converter is
described by a process of exchanging photons between two optical fields of
different frequencies. The Hamiltonian  representing such a system reads

${\displaystyle
\frac{\hat{H}}{\hbar }=\sum_{j=1}^{2}\omega _{j}\hat{a}_{j}^{\dagger }\hat{a}%
_{j}+i\lambda _{2}\left\{ \hat{a}_{1}^{\dagger }\hat{a}_{2}\exp [i(\omega
_{1}-\omega _{2})t]-{\rm h.c.}\right\}. }\hfill (1.2)$

\noindent This model can be applied to describe various optical phenomena, e.g. to
find analogies between frequency converter and beam splitter \cite{perin},
two-level atom driven by a single mode of electromagnetic field \cite{miel},
and Raman scattering \cite{perin,{orlov}}. The quantum properties of
parametric frequency converter are discussed in \cite{louis}. Furthermore,
some authors studied this model as the lossless linear coupler, e.g.
in \cite
{jansz}. In this situation the model is considered to be represented by two
electromagnetic waves which are guided inside the structure consisting of
two adjacent and parallel waveguides; the linear exchange of energy between
these two waveguides is established via the evanescent field \cite{marc}.

The derivation of the above Hamiltonians can be obtained from
the quantization of the cavity modes in a volume $V$. In  this case the total
energy of the field is given by \cite{louis,{add1}}

${\displaystyle
\hat{H}=}\frac{1}{8\pi }\int_{cavity}(
\epsilon
{\cal E} ^{2}{+}{\cal H}%
^{2})d^{3}V,\hfill (1.3a)$

\noindent where ${\cal E} $ and ${\cal H}$ are the electric and magnetic fields,
respectively, and $\epsilon $ is the dielectric constant.
To provide coupling between the various cavity modes we have to make the
dielectric constant $\epsilon $ varied according to

${\displaystyle
}\epsilon (R,t)=1+\Delta $ $\epsilon f(R)\sum_{i}\cos (\omega _{i}t+\phi
_{i}),\hfill (1.3b)$

\noindent where $\Delta \epsilon \ll 1$ and $f(R)$ is a function of the
position vector $R$, while $\phi _{i}$ is an arbitrary pump phase. From
the relations (1.3a) and (1.3b), and in terms of the boson operators,
 one can write the Hamiltonian (1.3a) in the following form

${\displaystyle
\frac{\hat{H}}{\hbar }=}\sum_{l}\omega _{l}(\hat{a}^{\dagger }\hat{a}+\frac{1%
}{2})-\sum_{i,l,m}k_{lm}\cos (\omega _{i}t+\phi _{i})(\hat{a}_{l}^{\dagger }-%
\hat{a}_{l})(\hat{a}_{m}^{\dagger }-\hat{a}_{m}),\hfill (1.3c)$

\noindent where $k_{lm}$ is the coupling coefficient having the form

${\displaystyle
}k_{lm}=\frac{\Delta \epsilon }{16\pi c^{2}}(\omega _{l}\omega _{m})^{\frac{1%
}{2}}\int_{cavity}f(r)u_{l}(r)u_{m}(r)dV,\hfill (1.3d)$

\noindent where the function $u_{l}(r)$  represents the normal mode and satisfies

${\displaystyle
\nabla \wedge \nabla \wedge }u_{l}(r)=(\frac{\omega _{l}}{c}%
)^{2}u_{l}(r).\hfill (1.3e)$

It is convenient  to refer to the recent work of Garrett et al. \cite{add2}, who
reported an experimental demonstration of phonon squeezing in a macroscopic
system by exciting a crystal KT$_{a}$O$_{3}$ with an ultrafast pulse of
light. The Hamiltonian relevant to their work can be adjusted to be
essentially equation (1.3c). Thus if one chooses the function $f(R)$ so that
$k_{lm}\neq 0$, this will leave an infinite number of modes coupled. Hence
by a proper choice of the pump frequency $\omega _{i}$, the interacting
modes will be limited to two modes, and then under certain conditions one
can obtain the Hamiltonians (1.1) or (1.2).

The optical processes involving the competition between
frequency converter and parametric amplifier are of interest from
theoretical and experimental points of view, which can be seen in three-mode
interaction \cite{mish1,{martin}} and nonlinear coupler \cite
{jansz2,{abdallf}}, where the nonlinear interaction couples different photon
modes and leads to the energy transfer between modes. Photons in some modes may
be annihilated, while those in the other modes created, and hence the photon
distribution is disturbed. In each time the rate of energy transfer between
the modes depends on the statistical properties of the light fields. This in
fact encouraged us to follow up the procedure given above and adjust the pump
frequency to extend the number of modes to be three instead of two. This
can lead to the following Hamiltonian

${\displaystyle
\frac{\hat{H}}{\hbar }=[\sum_{j=1}^{3}\omega _{j}\hat{a}_{j}^{\dagger }\hat{a%
}_{j}]+i\lambda _{1}\left\{ \hat{a}_{1}\hat{a}_{2}\exp [i(\omega _{1}+\omega
_{2})t]-{\rm h.c.}\right\} }\hfill $

${\displaystyle
+i\lambda _{2}\left\{ \hat{a}_{1}\hat{a}_{3}^{\dagger }\exp [i(\omega
_{1}-\omega _{3})t]-{\rm h.c.}\right\} +i\lambda _{3}\left\{ \hat{a}_{2}\hat{a}%
_{3}^{\dagger }\exp [i(\omega _{2}-\omega _{3})t]-{\rm h.c.}\right\} ,}\hfill
(1.4)$

\noindent which represents a model of three boson field modes, designated by
the annihilation operators $\hat{a}_{j},j=1,2,3$, and an intense
monochromatic light wave inducing parametric coupling between them. In this
Hamiltonian $\lambda _{j}$ are the coupling constants including the pump
amplitude and proportional to the second-order susceptibility of the medium $%
\chi ^{(2)}$ ; $\omega _{j}$ are the natural frequencies of oscillation of
the uncoupled modes and {\rm h.c.} is the Hermitian conjugate. This
interaction can also be established, e.g. by means of a bulk nonlinear
crystal exhibiting the second-order nonlinear properties in which three
dynamical modes of frequencies $\omega _{1},\omega _{2},\omega _{3}$ are
induced by three beams from lasers of these frequencies. When pumping this
crystal by means of the corresponding strong coherent pump beams, as
indicated in the Hamiltonian, we can approximately fulfil the phase-matching
conditions for the corresponding processes, in particular if the frequencies
are close each other (biaxial crystals may be helpful in such an
arrangement). Also a possible use of quasi-phase matching may help in the
realization, which is, however, more difficult technologically \cite{real}.
    There is another possibility to realize the interaction described by the
Hamiltonian (1.4) using a nonlinear asymmetric directional coupler
as shown schematically in Fig. 1.

For completeness, Hamiltonian (1.4) is a generalization for several models
used in the literature, e.g. \cite{moll1,{mish1},{bern1}}, and it is
invariant under the transformation $\hat{a}_{1}\longleftrightarrow \hat{a}%
_{2}$ with $\lambda _{2}\longleftrightarrow \lambda _{3}$. So that the
equations of the 2nd mode can be obtained from those of the first one.

\begin{picture}(70,70)(50,20)
\put (107,37){\vector(1,0){223}$\hat{a}_{1}(\frac{L}{v})$}

\put (107,60){\vector(1,0){223}$\hat{a}_{2}(\frac{L}{v})$}
\put (120,70){\framebox(0,0){$\hat{a}_{2}(0)$}}
\put (120,45){\framebox(0,0){$\hat{a}_{1}(0)$}}
\put (107,-20){\vector(1,0){223}$\hat{a}_{3}(\frac{L}{v})$}
\put (120,-10){\framebox(0,0){$\hat{a}_{3}(0)$}}

\put (300,-14){\vector(0,1){77}}
\put (200,0){\vector(0,1){40}}

\put (300,0){\vector(0,-1){20}}
\put (200,0){\vector(0,-1){20}}

\put (310,20){\framebox(0,0){$\lambda_{3}$}}
\put (210,20){\framebox(0,0){$\lambda_{2}$}}

\put (178,-50){\framebox(140,40){$\chi^{(1)}$}}
\put (178,30){\framebox(140,40){$\chi^{(2)}\quad\lambda_{1}$}}

\put (248,-100){\vector(1,0){70}}
\put (248,-100){\vector(-1,0){70}}
\put (248,-90){\framebox(0,0){$L$}}

\put (178,30){\line(0,-1){10}}
\put (178,10){\line(0,-1){10}}
\put (178,-10){\line(0,-1){10}}
\put (178,-30){\line(0,-1){10}}
\put (178,-50){\line(0,-1){10}}
\put (178,-70){\line(0,-1){10}}
\put (178,-90){\line(0,-1){10}}

\put (318,30){\line(0,-1){10}}
\put (318,10){\line(0,-1){10}}
\put (318,-10){\line(0,-1){10}}
\put (318,-30){\line(0,-1){10}}
\put (318,-50){\line(0,-1){10}}
\put (318,-70){\line(0,-1){10}}
\put (318,-90){\line(0,-1){10}}

\put (120,10){\line(1,0){20}}
\put (150,10){\line(1,0){20}}
\put (180,10){\line(1,0){20}}
\put (210,10){\line(1,0){20}}
\put (240,10){\line(1,0){20}}
\put (270,10){\line(1,0){20}}
\put (300,10){\line(1,0){20}}
\put (330,10){\line(1,0){20}}

\end{picture}

\vspace{1.7in}

{\it Fig.1 Scheme of realization of interaction in (1.4) using a nonlinear
asymmetric directional coupler which is composed of two optical waveguides
fabricated from first-order ($\chi ^{(1)}$) and second-order ($\chi ^{(2)}$)
materials, where $\chi $ designate susceptibility. Modes 1 and 2 propagate
in the first waveguide and mode 3 in the second waveguide. The interaction
between modes 1 and 2 is established by strong pump coherent light, which is
not indicated in figure, with the coupling constant $\lambda _{1}$. The
interactions between mode 3 and modes 1 and 2 are established linearly with
the coupling constants $\lambda _{2}$ and $\lambda _{3}$, respectively. The
beams are described by the photon annihilation operators as indicated; $z=vt$
is the interaction length and we assume that all beams have the same
velocity $v$ and the length of the waveguides is $L$. Outgoing fields are
examined as single or compound modes by means of homodyne detection to
observe squeezing of vacuum fluctuations, or by means of a set of
photodetectors to measure photon correlations, photon antibunching and
sub-Poissonian photon statistics in the standard ways.}
\vspace{.5cm}

In this article we investigate the statistical properties for the different
modes controlled by the Hamiltonian (1.4).  We consider the modes are initially
prepared in Fock (number) states
where these states represent the most basic quantum states and are maximally distant
from what one would call a classical field.
More illustratively, they can always exhibit sub-Poissonian statistics, however,
they provide no information on the phase since the number of photons is  quite certain.
Also any states can be expressed in terms of these
states with corresponding weighting distributions.
Further, there are various proposals how to generate such states.
For instance, these states  can be prepared in a nondemolition measurements
\cite{non1,{non2},{non3}} or in the micromaser \cite{non4}.
During  this work we use a finite number of quanta in each modes,
which is relevant to the experimental realization. Also we treat the model as
three harmonic oscillators interacting in a nonlinear crystal in a sense of
a competition between two frequency converters and one parametric
amplifier, as indicated in the Hamiltonian (1.4).
Under these circumstances the model is motivated by various interesting
results. For example,
we  show that this model can be used as an effective source for squeezed
light additionally to those systems given in the literature earlier \cite{squz}.
Also we show that it can yield thermal light and squeezed thermal
light.
Moreover, we demonstrate how the qualitative behaviour of the single
nonlinear process such as parametric amplifier or frequency converter can be
substantially changed in the presence of the other nonlinear processes.

The plan of the work is as follows:
In {\bf section 2} we introduce  the solution of
the equations of motion, {\bf section 3} is devoted to a
discussion of nonclassical phenomena such as quadrature squeezing,
second-order correlation function and the violation of Cauchy-Schwarz
inequality, in {\bf section 4} we demonstrate the quasidistribution
functions, and finally we summarize main conclusions in {\bf section 5}.

\section{ Solution of the equations of motion }

In this section we shall introduce the exact solution for the equations of
motion in the Heisenberg picture for the Hamiltonian (1.4).

The dynamics of the system is described by the Heisenberg equations of
motion which for any operator \^{O} are given by

${\displaystyle
\frac{d\hat{O}}{dt}=\frac{\partial \hat{O}}{\partial t}+\frac{1}{i\hbar }[%
\hat{O},\hat{H}],}\hfill (2.1)$

\noindent where [...,...] represents the commutator.

Therefore the propagation of the field operators, which are defined in the
slowly varying forms ($\hat{a}_{j}=\hat{A}_{j}\exp(-i\omega_{j}t),\quad
j=1,2,3$), can be written in the following form:

${\displaystyle
\frac{d\hat{A}_{1}}{dt}=-\lambda _{1}\hat{A}_{2}^{\dagger }-\lambda _{2}\hat{%
A}_{3}},\hfill (2.2a)$

${\displaystyle
\frac{d\hat{A}_{2}}{dt}=-\lambda _{1}\hat{A}_{1}^{\dagger }-\lambda _{3}\hat{%
A}_{3}},\hfill (2.2b)$

${\displaystyle
\frac{d\hat{A}_{3}}{dt}=\lambda _{2}\hat{A}_{1}+\lambda _{3}\hat{A}_{2}}%
.\hfill (2.2c)$

\noindent These are three equations with their Hermitian conjugates forming
a closed system which can be solved easily as

${\displaystyle
\hat{A}_{1}(t)=\hat{a}_{1}(0)f_{1}(t) +\hat{a}^{\dagger}_{1} (0)f_{2}(t)-%
\hat{a}_{2}(0)f_{3}(t) }\hfill $

${\displaystyle
-\hat{a}_{2}^{\dagger }(0)f_{4}(t)-\hat{a}_{3}(0)f_{5}(t)-\hat{a}%
_{3}^{\dagger }(0)f_{6}(t)},\hfill (2.3a)$

$\hfill $

${\displaystyle
\hat{A}_{2}(t)=\hat{a}_{2}(0)g_{1}(t) +\hat{a}^{\dagger}_{2} (0)g_{2}(t) -%
\hat{a}_{1}(0)g_{3}(t) }\hfill $

${\displaystyle
-\hat{a}_{1}^{\dagger }(0)g_{4}(t)-\hat{a}_{3}(0)g_{5}(t)-\hat{a}%
_{3}^{\dagger }(0)g_{6}(t)},\hfill (2.3b)$

$\hfill $

${\displaystyle
\hat{A}_{3}(t)=\hat{a}_{3}(0)h_{1}(t) +\hat{a}^{\dagger}_{3}(0)h_{2}(t) +%
\hat{a}_{2}(0)h_{3}(t)}\hfill $

${\displaystyle
+\hat{a}_{2}^{\dagger }(0)h_{4}(t)+\hat{a}_{1}(0)h_{5}(t)+\hat{a}%
_{1}^{\dagger }(0)h_{6}(t)},\hfill (2.3c)$

\noindent where the exact forms for the time-dependent coefficients, i.e.
for $f_{j}(t),g_{j}(t),h_{j}(t)$, which include all information
about the
structure of the model are complicated for the general case,
  we  give their form only for the  special case
when $\lambda_{2}=\lambda_{3}=\lambda$ (hence $f_{j}(t)=g_{j}(t)$)
, which will be frequently used
here:

${\displaystyle
f_{1,3}(t)= \pm\frac{1}{2} \left[ \cosh (\frac{\lambda_{1}t}{2})\cos (\bar{k}%
t) +\frac{\lambda_{1}}{2\bar{k}} \sinh (\frac{\lambda_{1}t}{2})\sin (\bar{k}%
t) \pm\cosh (\lambda_{1}t)\right],} \hfill $

${\displaystyle
f_{2,4}(t)= \mp\frac{1}{2} \left[ \sinh (\frac{\lambda_{1}t}{2})\cos (\bar{k}%
t) +\frac{\lambda_{1}}{2\bar{k}} \cosh (\frac{\lambda_{1}t}{2})\sin (\bar{k}%
t) \mp\sinh (\lambda_{1}t)\right],} \hfill $

${\displaystyle
f_{5}(t)=\frac{\lambda }{\bar{k}}\cosh (\frac{\lambda _{1}t}{2})\sin (\bar{k}%
t),\quad f_{6}(t)=-\frac{\lambda }{\bar{k}}\sinh (\frac{\lambda _{1}t}{2}%
)\sin (\bar{k}t),}\hfill (2.4)$

$\hfill$

${\displaystyle
h_{1}(t)= \cosh (\frac{\lambda_{1}t}{2})\cos (\bar{k}t) -\frac{\lambda_{1}}{2%
\bar{k}} \sinh (\frac{\lambda_{1}t}{2})\sin (\bar{k}t) ,} \hfill $

${\displaystyle
h_{2}(t)= - \sinh (\frac{\lambda_{1}t}{2})\cos (\bar{k}t) +\frac{\lambda_{1}%
}{2\bar{k}} \cosh (\frac{\lambda_{1}t}{2})\sin (\bar{k}t) ,} \hfill $

${\displaystyle
h_{3}(t)=h_{5}(t)=\frac{\lambda }{\bar{k}}\cosh (\frac{\lambda _{1}t}{2}%
)\sin (\bar{k}t),\quad h_{4}(t)=h_{6}(t)=-\frac{\lambda }{\bar{k}}\sinh (%
\frac{\lambda _{1}t}{2})\sin (\bar{k}t),}\hfill (2.5)$

\noindent where $\bar{k}=\sqrt{2\lambda^{2}-\frac{\lambda^{2}_{1}}{4}}$.
In fact, the nature of the solution can show how the interaction does work.
That is the time-dependent coefficients  contain both
trigonometric and hyperbolic functions.
Consequently, the energy associated with the propagating beams inside the
nonlinear crystal can be amplified as well as switched between
modes in the course of time.

On the basis of the well known commutation rules for boson operators, the
following relations can be proved between the time-dependent coefficients

${\displaystyle
f^{2}_{1}(t)-f^{2}_{2}(t)+f^{2}_{3}(t)-f^{2}_{4}(t)+f^{2}_{5}(t)
-f^{2}_{6}(t)=1,} \hfill $

${\displaystyle
f_{1}(t)g_{4}(t)
-f_{2}(t)g_{3}(t)+f_{3}(t)g_{2}(t)-f_{4}(t)g_{1}(t)-f_{5}(t)g_{6}(t)
+f_{6}(t)g_{5}(t)=0,} \hfill $

${\displaystyle
f_{1}(t)g_{3}(t)-f_{2}(t)g_{4}(t)+f_{3}(t)g_{1}(t)-f_{4}(t)g_{2}(t)-f_{5}(t)g_{5}(t)+f_{6}(t)g_{6}(t)=0.%
}\hfill (2.6)$

\noindent The remain relations can be obtained from (2.6) by means of the
following transformations

${\displaystyle
\Bigl(f_{1}(t),f_{2}(t),f_{3}(t) ,f_{4}(t),f_{5}(t),f_{6}(t)\Bigr) } \hfill $

${\displaystyle
\longleftrightarrow \Bigl(-g_{3}(t),-g_{4}(t),-g_{1}(t)
,-g_{2}(t),g_{5}(t),g_{6}(t)\Bigr) } \hfill $

${\displaystyle
\longleftrightarrow \Bigl%
(h_{5}(t),h_{6}(t),-h_{3}(t),-h_{4}(t),-h_{1}(t),-h_{2}(t)\Bigr).}\hfill
(2.7)$

Based on the results of the present section, we can study the
quantum properties of the evolution of the different modes in the model when
they are initially prepared in Fock states.

\section{ Nonclassical phenomena}

Since   the photons produced in a nonlinear optical process are known to
possess unusual correlation properties resulting in many of nonclassical
aspects of the radiation field. Therefore, we investigate some of these
properties for the model under consideration in the present section. Our
attention is focused on squeezing phenomenon,  sub-Poissonian photon statistics
and the violation  of Cauchy-Schwarz inequality.

\subsection{Squeezing phenomenon}

The concept of squeezing of a quantum electromagnetic field has been given
a great deal of interest in the view of the possibility of reducing the noise
of an optical signal below the vacuum limit and the possible potential
application in optical communication networks \cite{yun2}
and gravitational wave detection \cite{cav1}.
This light can be measured by homodyne detection where the signal is
superimposed on a strong coherent beam of the local oscillator.
The generation of such light has  been reported in  various schemes \cite{sq}.

Generally there are several definitions for squeezing, e.g. see \cite{squz}.
Here  we investigate some of these types for the model under discussion,
in particular, single-mode, two-mode and three-mode squeezing as well as
sum-squeezing \cite{hil} when the modes are prepared initially in Fock states.
The starting point for this study are
 the two quadratures $\hat{X}$ and $\hat{Y}$
which are related to the conjugate electric and
magnetic field operators $\hat{E}$ and $\hat{H}$. They are defined in the
standard way. Assuming that these two quadrature operators satisfy the
following commutation relation

${\displaystyle
\left[ \hat{X},\hat{Y}\right] =\hat{C},}\hfill (3.1a)$

\noindent where $\hat{C}$ may be an operator or a c-number with respect to
which kind of squeezing we consider, the following uncertainty
relation holds

${\displaystyle
\langle (\triangle \hat{X})^{2}\rangle \langle (\triangle \hat{Y}
)^{2}\rangle \geq \frac{|\langle \hat{C}\rangle |^{2}}{4},}\hfill (3.1b)$

\noindent where $\langle (\triangle \hat{X})^{2}\rangle=\langle \hat{X}
^{2}\rangle -\langle \hat{X}\rangle ^{2}$ is the variance.
Therefore, we can say that the model possesses $X$-quadrature squeezing
if the $S$-factor \cite{mad},

${\displaystyle
S(t)=\frac{\langle (\triangle \hat{X}(t))^2\rangle - 0.5|\langle \hat{C}\rangle | }
{0.5|\langle \hat{C}\rangle |}
,}\hfill (3.1c)$

\noindent satisfies the inequality $-1\leq S<0$. Similar expression for
the $Y$-quadrature ($Q$-factor) can be obtained.

Firstly, for three-mode squeezing we define the two quadratures as

${\displaystyle \hat{X}_{3}(t)=\frac{1}{2}[ \hat{A}_{1}(t)+\hat{A}_{2}(t) +
\hat{A}_{3}(t)+\hat{A}_{1}^{\dagger}(t)+\hat{A}_{2}^{\dagger}(t)+ \hat{A}%
_{3}^{\dagger}(t)],}\hfill (3.2a)$

${\displaystyle \hat{Y}_{3}(t)=\frac{1}{2i}[ \hat{A}_{1}(t)+\hat{A}_{2}(t)+
\hat{A}_{3}(t)-\hat{A}_{1}^{\dagger}(t)-\hat{A}_{2}^{\dagger}(t)- \hat{A}%
_{3}^{\dagger}(t)],}\hfill (3.2b)$

\noindent
where the subscript $3$ in the left-hand side stands for
three mode case.
Using (2.3) together with the definition of the variance, for initial
Fock states  we arrive at

${\displaystyle
4\langle (\triangle \hat{X}_{3}(t))^{2}\rangle= (1+2\bar{n}%
_{1})[f_{1}(t)+f_{2}(t)+h_{5}(t)+h_{6}(t)-g_{3}(t) -g_{4}(t)]^{2} }\hfill $

${\displaystyle
+(1+2\bar{n}_{2})[f_{3}(t)+f_{4}(t)-g_{1}(t)-g_{2}(t)
-h_{3}(t)-h_{4}(t)]^{2} }\hfill$

${\displaystyle
+(1+2\bar{n}_{3})[f_{5}(t)+f_{6}(t)+g_{5}(t)+g_{6}(t)-h_{1}(t)
-h_{2}(t)]^{2}, }\hfill (3.3a)$

$\hfill $

${\displaystyle
4\langle (\triangle \hat{Y}_{3}(t))^{2}\rangle= (1+2\bar{n}%
_{1})[f_{1}(t)+g_{4}(t)+h_{5}(t)-f_{2}(t)- g_{3}(t)-h_{6}(t)]^{2} }\hfill $

${\displaystyle
+(1+2\bar{n}_{2})[f_{3}(t)+g_{2}(t)+h_{4}(t)-f_{4}(t)-g_{1}(t) -h_{3}(t)]^{2}%
}\hfill$

${\displaystyle
+(1+2\bar{n}_{3})[f_{5}(t)+g_{5}(t)+h_{2}(t)-f_{6}(t)-g_{6}(t)
-h_{1}(t)]^{2}; }\hfill (3.3b)$

\noindent
$\bar{n}_{j}$ is the mean photon number for the $j$th
mode. The expressions for the single-mode and two-mode squeezing can be
obtained easily from (3.3) for two quadratures defined in a manner analogous
to (3.2) by dropping the coefficients of the absent mode, e.g. for the 1st
mode, single-mode squeezing can be obtained by setting $g_{j}(t)=h_{j}(t)= 0,
j=1,2,3$
in (3.3). It should be taken into account that $\hat{C}$ is a
c-number in this case and equals $\frac{1}{2},1,\frac{3}{2}$
corresponding to the single-mode, two-mode and three-mode squeezing,
respectively.

We start our discussion when
 $\lambda_{2}=\lambda_{3}=\lambda$.
In this case  the quadrature variances
for the 3rd mode (single-mode squeezing) are

${\displaystyle
\langle (\triangle \hat{X}_{1}(t))^{2}\rangle=
\frac{1}{4}\left\{
2(1+\bar{n}_{1}+\bar{n}_{2})
[ \frac{\lambda}{\bar{k}}\sin (\bar{k}t)]^{2} \right.
}\hfill $

${\displaystyle
+ \left.
(1+2\bar{n}_{3})[ \cos (\bar{k}t) +\frac{\lambda_{1}}{2\bar{k}}
\sin (\bar{k}t)]^{2}\right\}\exp (-\lambda_{1}t), }\hfill (3.4a)$

$\hfill$

${\displaystyle
\langle (\triangle \hat{Y}_{1}(t))^{2}\rangle=
\frac{1}{4}\left\{
2(1+\bar{n}_{1}+\bar{n}_{2})
[ \frac{\lambda}{\bar{k}}\sin (\bar{k}t)]^{2} \right. }\hfill $

${\displaystyle
+\left.
(1+2\bar{n}_{3})
[ \cos (\bar{k}t) -\frac{\lambda_{1}}{2\bar{k}}\sin (\bar{k}t)]^{2}
\right\}\exp (\lambda_{1}t)
. }\hfill (3.4b)$

\noindent
The immediate conclusion can be drawn from these  expressions which is  that
there is squeezing in the $X$-quadrature after switching on of the
interaction by a suitable time as a result of the fact that the Fock
states are not minimum-uncertainty states.
Indeed, this is an interesting result because most of the
nonlinear optical processes including correlation of modes are unable to
provide such a property.
Furthermore, under this condition
similar expressions, i.e. $(...)\exp (-\lambda_{1}t)$ for $X$-quadrature
and $(...)\exp (\lambda_{1}t)$ for $Y$-quadrature,
 can be obtained
for the  two-mode squeezing, in particular, between the 1st and 2nd modes
and also for the three-mode squeezing.
Also, this situation is still valid if one considers that the modes
are initially in the states which are represented by a
density matrix  diagonal in the number state basis (thermofield states);
 in this case
$\bar{n}_{j}$  represents the average photon number for the $j$th thermal
mode and therefore the behaviour of the model under consideration
is close to that of  a microwave Josephson-junction parametric amplifier
\cite{yur}, in which a thermal input field has been introduced to
the squeezing device and the generated field has exhibited noise reduction.
Indeed, thermal noise  is inevitable and hard to quench,
so that it is more realistic to  consider a thermal state instead of
a vacuum state input to a squeezing device \cite{paul1}.
More  details about
the properties of squeezed number and  squeezed thermal states
can be found in \cite{ paul1,{kim3}}.
We proceed by  plotting the squeezing factors associated with the
$X$-quadrature, i.e. $S$-factor, for the three types
when the  modes are initially in the Fock states as shown
in Fig. 2, for the given values of the parameters.
In this figure we have considered the single-mode (star-centered curve-3rd
mode), two-mode (short-dashed curve-(1,2) modes) and three-mode squeezing
(long-dashed curve).  For the sake of comparison  we adopted
solid curve for squeezing factor of well-known squeezed
number states \cite{fear}, i.e. for $S_{1}(t)=(1+2\bar{n})
\exp(-\lambda_{1}t)-1, \quad r=\lambda_{1}t$ is the usual squeeze parameter.
From this figure
we can see that for all cases at the first moments of the evolution
there is no squeezing owing to the nature of  the Fock states.
Further, the three types have approximately similar behaviours
in a sense that they oscillate  around  the value
 of the usual squeezed number states, i.e. around the solid  curve, as a result of
 switching of the energy between the interacting modes in the system.
The significant remark is that they arrive to the steady state
(maximum squeezing)  at the large values of interaction times.

From the above analysis it is clear that the 3rd mode possesses  more
pronounced nonclassical behaviour compared with the 1st and 2nd modes.
This is related to the structure of the Hamiltonian (1.4) and can be
interpreted as follows: modes 1 and 2 undergo amplification from the
first process (i.e. from the parametric amplifier)
and then the energy transfers from these two modes to the 3rd mode via
the  converter processes.
Indeed, such a structure is the source of the single-mode squeezing
   which has not been seen before for
the most of the important models involving three-mode interaction process,
e.g. \cite{mish1,{abdal}}.

\setcounter{figure}{1}
\begin{figure}[h]%
    \includegraphics[width=8cm]{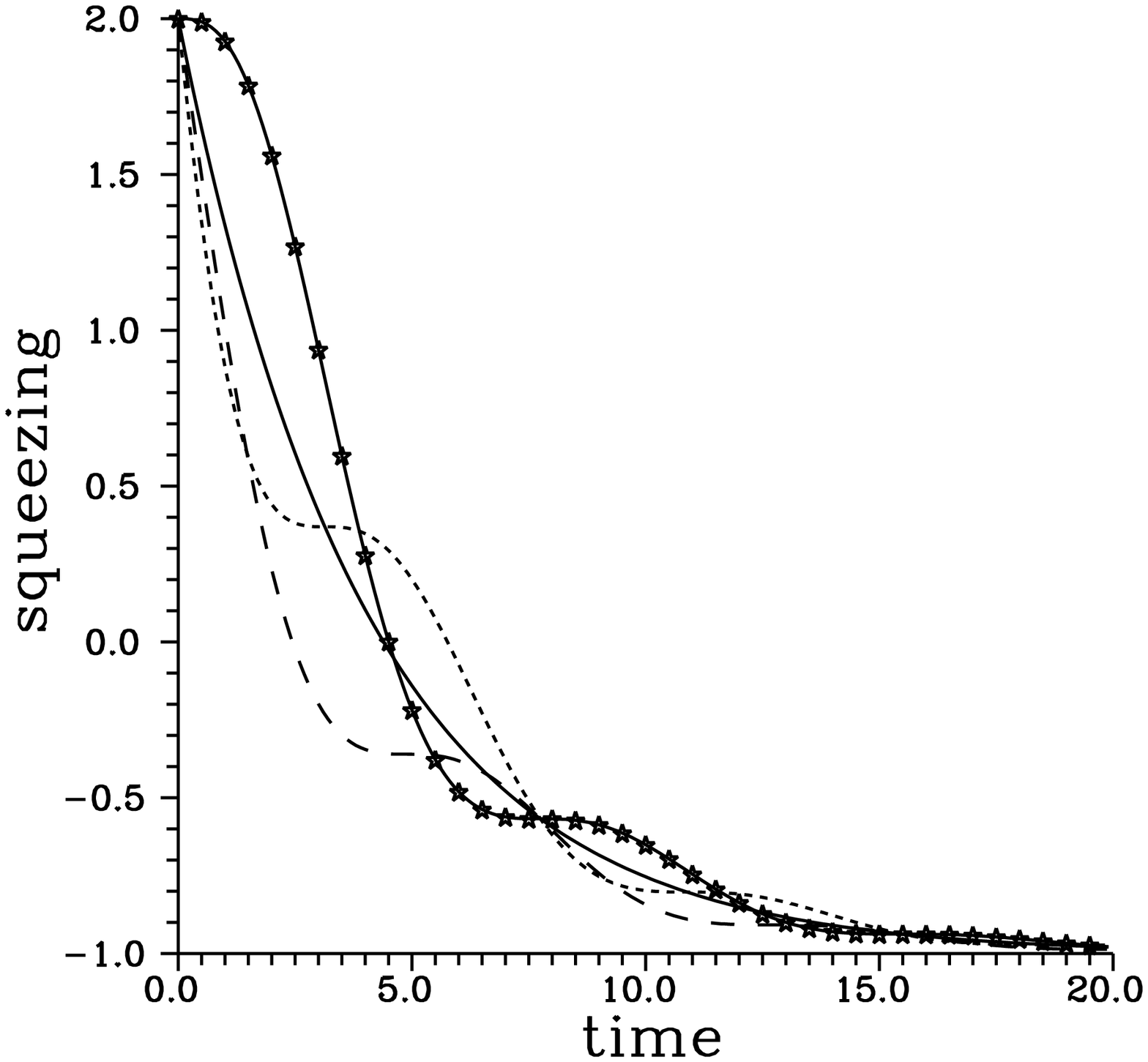}
   \caption{
Squeezing factor, $S_{n}(t)$, for initial number states with mean
photon numbers $\bar{n}_{j}=1,\quad j=1,2,3$, against the time for
the single-mode (star-centered curve-3rd mode), two-mode
(short-dashed curve-(1,2) modes) and three-mode squeezing
(long-dashed curve) for $(\lambda_{1},\lambda_{2}
,\lambda_{3})=(0.25,0.3,0.3)$; the solid curve is for squeezed
number states, i.e. for $S_{1}(t)=(1+2n)\exp(-\lambda_{1}t)-1,
\quad \lambda_{1}t$ represents the usual squeeze parameter $r$.  }
  \label{fig2}
\end{figure}

Secondly, we discuss sum-squeezing \cite{hil} for the model under discussion.
Sum-squeezing effect is both higher-order and multimode phenomenon
and the states join to this class are nonclassical states since its Glauber-Sudarshan
$P$-function  always does not exist and it is not  well defined  function.
Also, there is a connection between sum-squeezing  and sum-frequency generation
and consequently it can be converted into normal single-mode squeezing by an
appropriate nonlinear optical process and this may be used for the detection.
The importance of this  type is related to
the fact that squeezing can  exist for the correlated modes even if the
individual modes are not squeezed in the normal sense.

Now  the quadratures corresponding to the real and imaginary parts for
sum-squeezing are given by \cite{hil}

${\displaystyle \hat{X}_{s}(t)=\frac{1}{2}[ \hat{A}_{j}(t)\hat{A}_{k}(t) +
\hat{A}_{j}^{\dagger}(t)\hat{A}_{k}^{\dagger}(t)
],}\hfill (3.5a)$

${\displaystyle
 \hat{Y}_{s}(t)=\frac{1}{2i}[ \hat{A}_{j}(t)\hat{A}_{k}(t)
-\hat{A}_{j}^{\dagger}(t)\hat{A}_{k}^{\dagger}(t)
],}\hfill (3.5b)$

\noindent where $j\neq k$ and the subscript $s$ stands for sum-squeezing.
In this case $\hat{C}= \hat{A}_{j}^{\dagger}(t)\hat{A}_{j}(t) +
\hat{A}_{k}^{\dagger}(t) \hat{A}_{k}(t) +1$, i.e. it is a state dependent.
One should notice that the amplitude squared squeezing \cite{add3}
is the degenerate limit of sum-squeezing, and also the operators
$\hat{X}_{s}(t),\hat{Y}_{s}(t),\hat{C}$ form a representation of the
$su(1,1)$ Lie algebra and thus they have been used in
the interferometric measurements \cite{yurr}.

\begin{figure}[h]%
    \includegraphics[width=8cm]{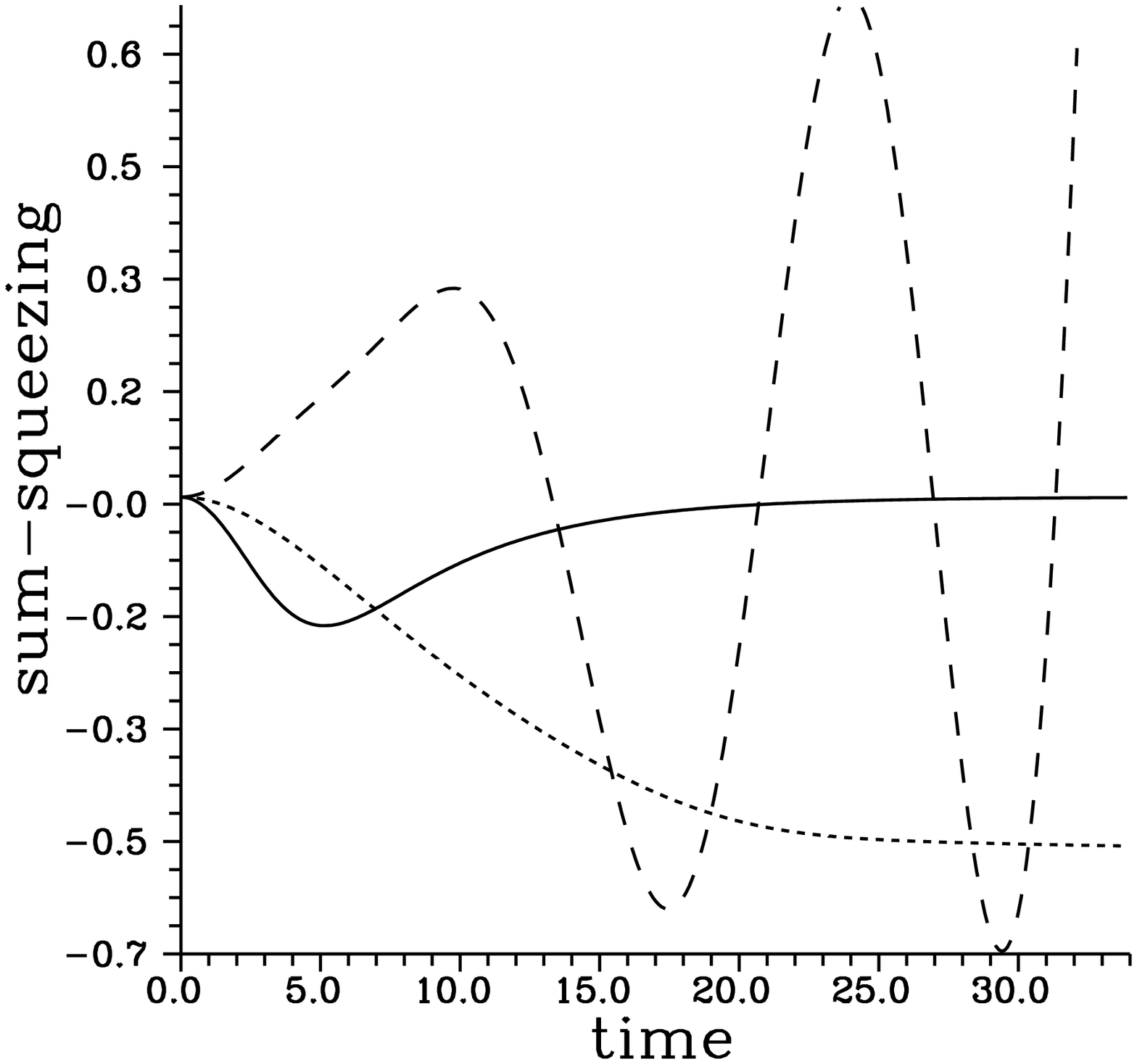}
   \caption{
Sum-squeezing factor for initial number states with mean photon
numbers $\bar{n}_{j}=1,\quad j=1,2,3$, against the time  for
$\lambda_{j}=0.1, j=1,2,3$ and for the single parametric amplifier
($Y$-component, solid curve), between modes 1,2 ($Y$-component,
short-dashed curve) and between modes 1,3 ($X$-component,
long-dashed curve).  }
  \label{fig3}
\end{figure}

The expressions for $S(t)$ and $Q(t)$  for sum-squeezing  are
rather lengthy in this case, further, they can be calculated straightforwardly and are not
very illuminating, so we shall not quote their explicit forms.
As  is known the single parametric amplifier is the source of
sum-squeezed light \cite{hils}, however, single frequency converter cannot
produce such a light. The system under consideration can produce sum-squeezed
light which is more effective than that of the single parametric amplifier
and this reflects the role of competition of different processes in the model.
These facts are clear in Fig. 3 where the sum-squeezing factor is plotted
against the time for the shown values of the parameters.
Solid, short-dashed and long-dashed curves show that sum-squeezing
factor corresponds to the $Y$-component of single parametric amplifier,
the $Y$-component  of (1,2) compound mode and the $X$-component
of (1,3) compound mode, respectively.
From this figure one can observe that for the single parametric amplifier
(solid curve), it starts from squeezing bound, goes to its maximum squeezing
value ($\simeq 12\%$) and then squeezing values decrease and eventually comes back to the
 initial values for large interaction
times. However, for (1,2) mode, squeezing appears after switching on
of
the interaction as before, then its value increases in the course of time
and eventually locks to give $50\%$ squeezing in the large time of interaction
domain. It is important to point out neither the 1st mode nor the 2nd one
is squeezed in the normal sense in this case.
We proceed by discussing the quantity for   (1,3) mode (long-dashed
curve). One can observe that squeezing is periodically established displaying
its maximum value in the large interaction
time and then squeezing disappears (this is not indicated in the figure).
The maximum value of the sum-squeezing in this case is more pronounced
than in the previous ones, in particular, up to $70\%$ squeezing can be
obtained.
We should remind here that the 3rd mode can produce squeezing
in the normal sense (cf. (3.4)).  Consequently we can conclude that there
is no connection between sum-squeezing and normal squeezing for the
models including correlation between modes \cite{hil}.
It is important to point out that the behaviour of the
sum-squeezing in $(2,3)$ mode
 is similar to that for $(1,3)$ mode
 and this  can be recognized from the Hamiltonian (1.4).
Final remark, the squeezing values are sensitive to the initial mean
photon-numbers of the modes which can vanish  completely  for its large values.

\subsection{ Sub-Poissonian statistics }

Sub-Poissonian light is an example of nonclassical light which can be
measured by photodetectors. A state (of a single mode for convenience) which
displays sub-Poisson statistics is characterized by the fact that the
variance of the photon number $\langle (\triangle \hat{n}_{j}(t))^{2}\rangle$
is less than the average photon number $\langle \hat{n}_{j}(t)\rangle
=\langle \hat{A}^{\dagger}_{j}(t) \hat{A}_{j}(t)\rangle$. This can be
expressed by means of the normalized normal second-order correlation function
as

${\displaystyle
g_{j}^{(2)}(t)=\frac{\langle \hat{A}_{j}^{\dagger 2}(t)\hat{A}
_{j}^{2}(t)\rangle }{\langle \hat{A}_{j}^{\dagger }(t)\hat{A}_{j}(t)\rangle
^{2}}}$

$\hfill $

${\displaystyle
=1+\frac{\langle (\triangle \hat{n}_{j}(t))^{2}\rangle -\langle \hat{A}%
_{j}^{\dagger }(t)\hat{A}_{j}(t)\rangle }{\langle \hat{A}_{j}^{\dagger }(t)%
\hat{A}_{j}(t)\rangle ^{2}},}\hfill (3.6)$

\noindent where the subscript $j$ relates to the $j$th mode. Then it holds
that $g_{j}^{(2)}(t)<1$ for sub-Poissonian distribution, $g_{j}^{(2)}(t)>1$
for super-Poissonian distribution and when $g_{j}^{(2)}(t)=1$ Poisson
distribution of coherent photons occurs.
Also the system exhibits maximum sub-Poissonian statistics when
$g_{j}^{(2)}(t)=0$.
Here we may mention that Fock state,
chaotic field and coherent state are representing good
examples for
sub-Poissonian, super-Poissonian and Poissonian statistics, respectively.
Moreover, the generation of sub-Poissonian light has been established in
a semiconductor laser \cite{yama} and in the microwave region using masers
operating in the microscopic regime \cite{rem} (for review see
\cite{progress}).

On the other hand, it is worth referring to photon antibunching (bunching)
phenomenon, which is generally   not
equivalent to   sub-Poissonian (super-Poissonian) photon
statistics \cite{singh,{zou}}.
The basic formula to study this phenomenon
is the two-time normalized intensity  correlation function \cite{zou,
{miranowicz}}, which can be  represented as
the joint detection probability of  two photons, one at time $t$
and another at time $t+\tau$, where $\tau$ is the time interval between
the two-photon detection process. Nevertheless,
 the bunching/antibunching and super-/sub-Poissonian statistics
 are identical to the stationary fields.
For more details about the connection between these two phenomena, different
definitions as well as  applications the reader can consult
\cite{miranowicz}.

Now we investigate the sub-Poissonian statistics for our model when all
modes are initially in the Fock states as we did for squeezing.
In this case
 the moments $\langle \hat{A}^{\dagger}_{1}(t) \hat{A}_{1}(t)
\rangle$ and $\langle \hat{A}^{\dagger 2}_{1}(t) \hat{A}^{2}_{1}(t) \rangle$
for the 1st mode read

${\displaystyle
\langle \hat{A}^{\dagger}_{1}(t) \hat{A}_{1}(t) \rangle
=[f^{2}_{1}(t)+f^{2}_{2}(t)]\bar{n}_{1}+ [f^{2}_{3}(t)+f^{2}_{4}(t)]\bar{n}%
_{2}}\hfill $

${\displaystyle
+[f_{5}^{2}(t)+f_{6}^{2}(t)]\bar{n}%
_{3}+f_{2}^{2}(t)+f_{4}^{2}(t)+f_{6}^{2}(t),}\hfill (3.7a)$

$\hfill$

${\displaystyle
\langle \hat{A}^{\dagger 2}_{1}(t) \hat{A}^{2}_{1}(t) \rangle =f^{4}_{1}(t)%
\bar{n}_{1}( \bar{n}_{1} -1) + f^{4}_{2}(t)(\bar{n}_{1}+1)( \bar{n}_{1} +2)
+ f^{4}_{3}(t)\bar{n}_{2}( \bar{n}_{2} -1) }\hfill $

${\displaystyle
+f^{4}_{4}(t)(\bar{n}_{2}+1)( \bar{n}_{2}+2) +f^{4}_{5}(t)\bar{n}_{3}( \bar{n%
}_{3} -1) +f^{4}_{6}(t)(\bar{n}_{3}+1)( \bar{n}_{3} +2) }\hfill$

${\displaystyle
+ f^{2}_{5}(t)f^{2}_{6}(t)(2\bar{n}_{3}+1)^{2} +[f_{1}(t)f_{2}(t)(2\bar{n}%
_{1}+1)+ f_{3}(t)f_{4}(t)(2\bar{n}_{2}+1)]^{2} }\hfill$

${\displaystyle
+4f^{2}_{1}(t)\bar{n}_{1} [f^{2}_{3}(t) \bar{n}_{2}+f^{2}_{4}(t)(\bar{n}%
_{2}+1)]}\hfill $

${\displaystyle
+4f^{2}_{2}(t)[f^{2}_{3}(t)\bar{n}_{2} (\bar{n}_{1}+1)+f^{2}_{4}(t)(\bar{n}%
_{1}+1) (\bar{n}_{2}+1)] }\hfill$

${\displaystyle
+2(2\bar{n}_{3}+1)f_{5}(t)f_{6}(t)[f_{1}(t)f_{2}(t)(2\bar{n}_{1}+1)+
f_{3}(t)f_{4}(t)(2\bar{n}_{2}+1)]}\hfill $

${\displaystyle
+4[f_{5}^{2}(t)\bar{n}_{3}+f_{6}^{2}(t)(\bar{n}_{3}+1)][f_{1}^{2}(t)\bar{n}%
_{1}+f_{2}^{2}(t)(\bar{n}_{1}+1)+f_{3}^{2}(t)\bar{n}_{2}+f_{4}^{2}(t)(\bar{n}%
_{2}+1)];}\hfill (3.7b)$

\noindent as we have mentioned earlier $\bar{n}_{j}$ is the mean photon
number of the $j$th mode.
 The expressions associated with the 2nd and 3rd modes
can be obtained from (3.7) using the transformations (2.7).

Let us start our discussion  when  $\bar{n}_{1}=\bar{n}_{2}=\bar{n}_{3}=\bar{n}$
and also consider the case of negligible contribution from parametric amplifier ($\lambda _{1}\sim
0$). In this case the normalized normal second-order correlation
function for the 1st mode reduces to

${\displaystyle
g_{1}^{(2)}(t)=2-(1+\frac{1}{\bar{n}}%
)[f_{1}^{4}(t)+f_{3}^{4}(t)+f_{5}^{4}(t)].}\hfill (3.8)$

\noindent From the identity (2.6) one can show that $
0<(f^{4}_{1}(t)+f^{4}_{3}(t)+f^{4}_{5}(t))\leq 1$ and thus

${\displaystyle
2>g_{1}^{(2)}(t)\geq (1-\frac{1}{\bar{n}}),}\hfill (3.9)$

\noindent
 which means that  the model, in this case, can exhibit only sub-Poissonian
statistics or partial coherence.

\begin{figure}[h]%
  \centering
  \subfigure[]{\includegraphics[width=8cm]{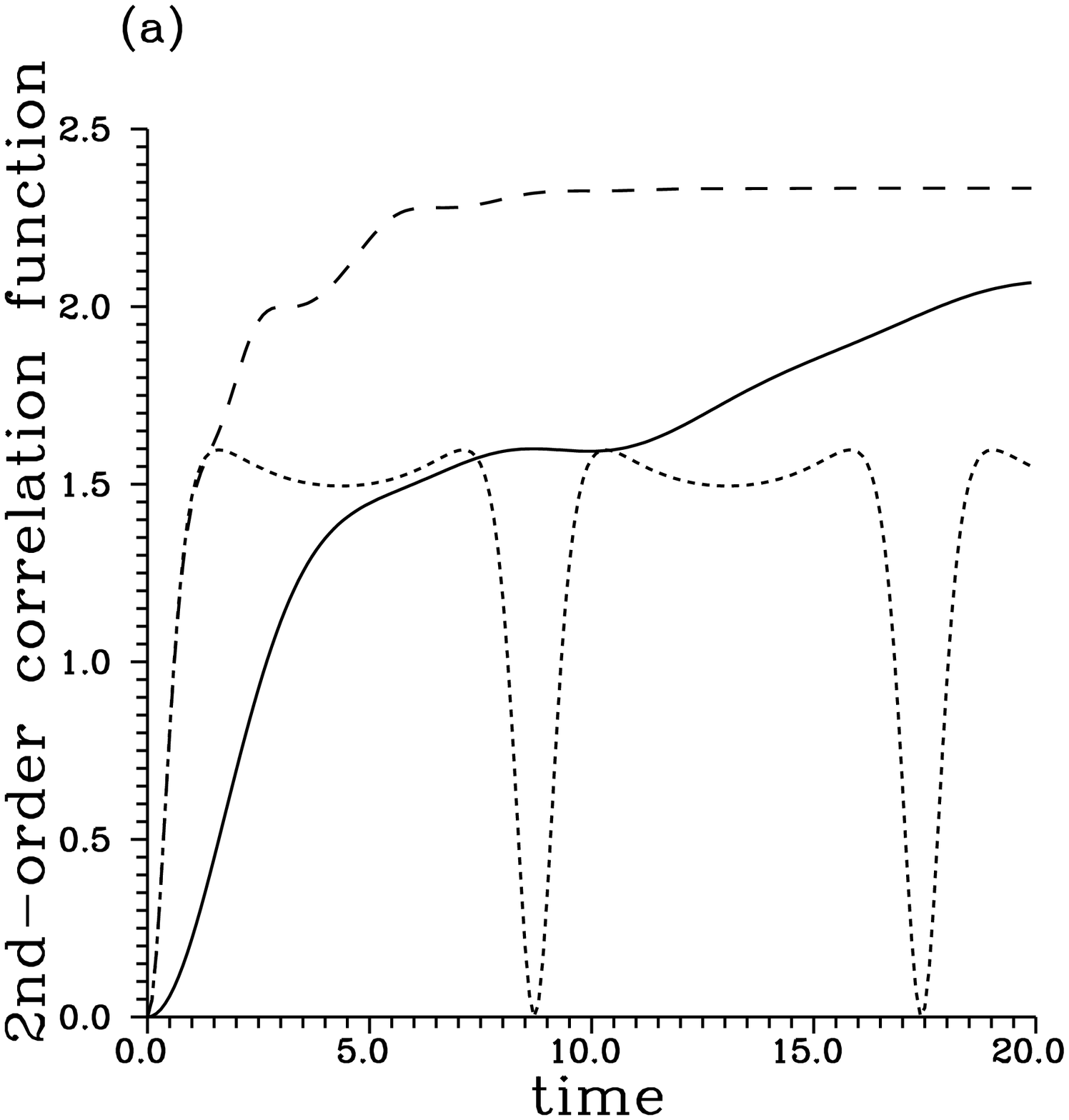}}
 \subfigure[]{\includegraphics[width=8cm]{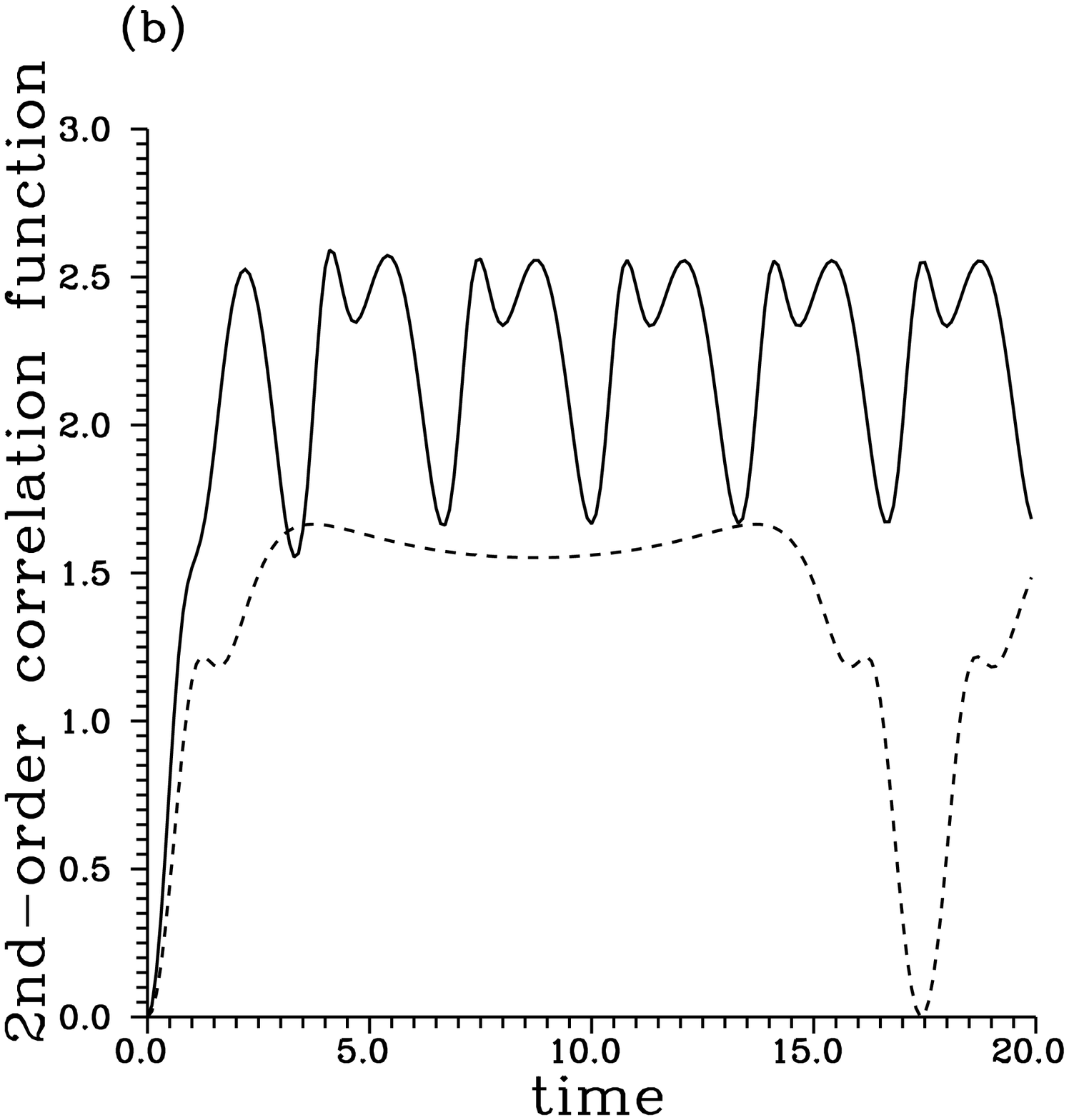}}
    \caption{
Normalized normal second-order correlation function
$g_{j}^{(2)}(t)$ against time for the modes which are initially in
the number state $|1,1,1\rangle$ for a) the 1st mode   for
$(\lambda_{1},\lambda_{2},\lambda_{3})=(0.6,0.7,0)$ (short-dashed
curve), $(0.6,0.7,0.7)$ (long-dashed curve), and $(0.1,0.2,0.2)$
(solid curve);
 b) the 3rd mode  for $(\lambda_{1},\lambda_{2},\lambda_{3})=(0.6,0.7,0)$
 (dashed curve), and $(0.6,0.7,0.7)$ (solid curve).  }
  \label{fig4}
\end{figure}

Now we can investigate  the evolution of $g_{j}^{(2)}(t)$ when the modes
are initially prepared in the number state $|1,1,1\rangle $ for the 1st
and 3rd modes, as shown in Figs. 4a,b, respectively, for shown values.
In Fig. 4a one can observe that $g_{1}^{(2)}(t)$ has smoothed behaviour
(see, solid and
long-dashed curves), i.e. it goes rapidly from values corresponding to the
sub-Poissonian statistics to those for super-Poissonian ones and becomes
almost stable for large interaction times. We further see that the stronger
the interaction, the shorter the sub-Poissonian interval. Similar behaviour
is available in a short interaction time if one turns his attention to $
g_{3}^{(2)}(t)$, see Fig. 4b. However, in the large interaction times $
g_{3}^{(2)}(t)$ becomes stable in a sense of periodic oscillations.
So, generally, we can see that for the Hamiltonian model (1.4), the modes lose their
sub-Poissonian character and provide partially coherent, chaotic and
superchaotic statistics for large interaction time.
We should comment here  that the behaviour of the 3rd mode in this
model is different
from that of the usual squeezed  number states given in the literature
earlier \cite{yun2} even if they can provide a similar behaviour for the quadrature
squeezing. This fact is clear by comparing the behaviour of the second-order
correlation function of the 3rd mode with that of the usual number states
\cite{kim3}
where for the latter (with $\bar{n}\neq 0$) $g^{(2)}(0)$ has smoothed behaviour and satisfies the
inequality $0\leq g^{(2)}(0)\leq 1.7$.  On the other hand, if one
compares the short-dashed curves with the solid ones in Figs. 4a,b, it is
easy to recognize that the two-mechanism interaction system \cite{mish1} ($%
\lambda _{3}=0$) conserves the initial sub-Poissonian
statistics of the interacting modes which can be periodically recovered with
period $t\simeq 8$ (for the 1st mode) and $t\simeq 16$ (for the 3rd mode).
Also similar approach is available for the model including competition
between two parametric amplifiers and one parametric frequency converter \cite
{abdal}.

\begin{figure}[h]%
    \includegraphics[width=8cm]{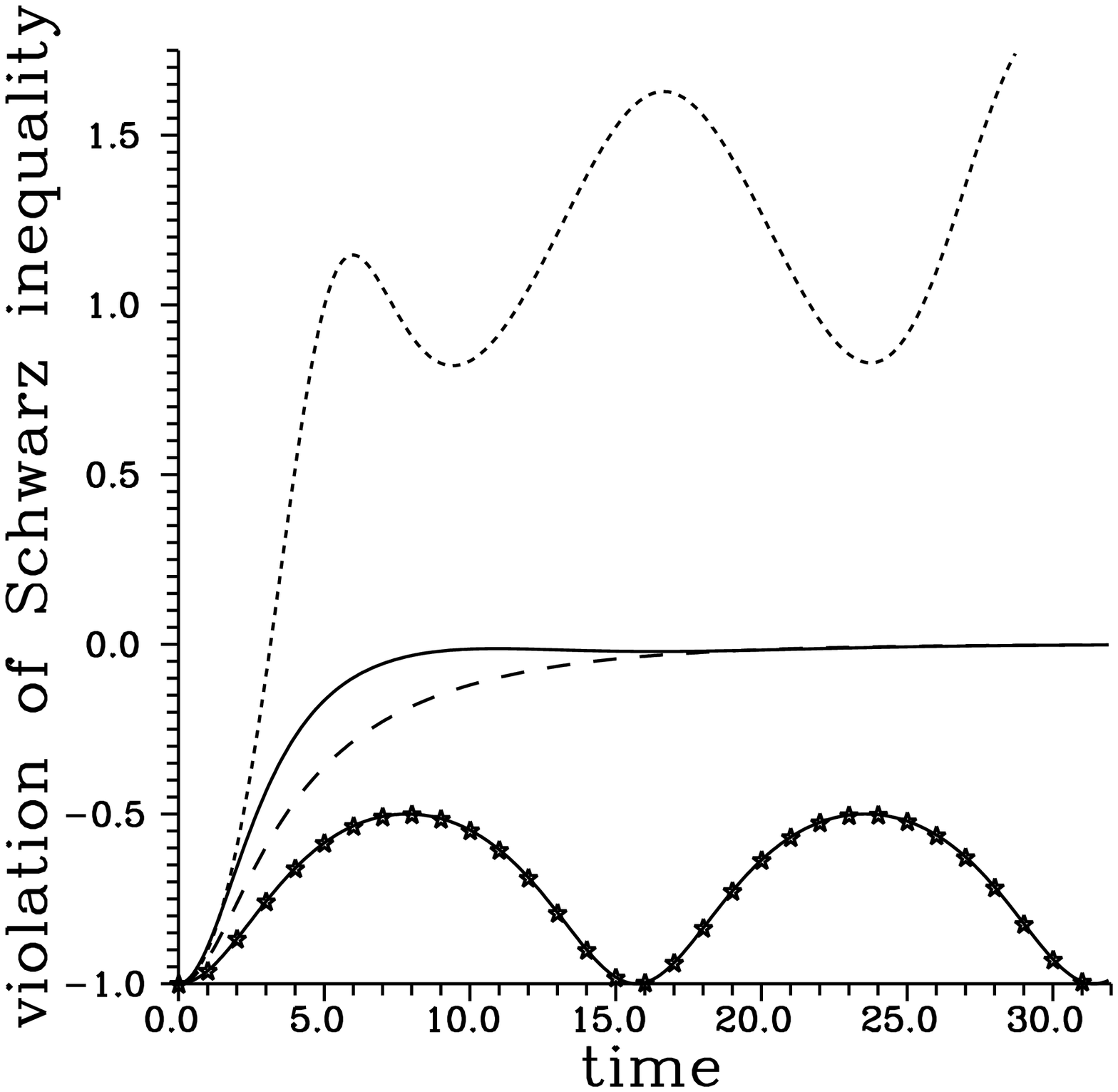}
   \caption{
Evolution of the quantity $I_{j,k}(t)$
 against the time for $(\bar{n}_{j},\lambda_{j})=(1,0.1), j=1,2,3$.
 The solid and  short-dashed curves represent
the above quantity obtained in $(1,2)$ and $(1,3)$ modes,
respectively. The long-dashed  and star-centered curves are
 given for $(\lambda_{1},\lambda_{2},\lambda_{3})=(0.1,0,0)$ and $(0,0.1,0)$
, respectively, with $\bar{n}_{j}=1$ .
  }
  \label{fig5}
\end{figure}

The second quantity  we are planning to study  here is
the violation of Cauchy-Schwarz inequality. This quantity shows the
anticorrelation  effects in the compound modes and  can be
observed in a two-photon interference experiment \cite{ghos}.
This quantity  can be
represented by the factor \cite{ag1}

${\displaystyle
I_{j,k}(t)=\frac{[\langle \hat{A}^{\dagger 2}_{j}(t)\hat{A}^{2}_{j}(t) \rangle
\langle
\hat{A}^{\dagger 2}_{k}(t)\hat{A}^{2}_{k}(t)\rangle]^{\frac{1}{2}}}
{ \langle \hat{A}^{\dagger}_{j}(t)\hat{A}_{j}(t)\hat{A}^{\dagger}_{k}(t)
 \hat{A}_{k}(t)\rangle }-1. }
\hfill (3.10) $

\noindent The negative values for the quantity $I_{j,k}(t)$ mean that the
intermodal correlation is larger than the correlation between  photons in
the same mode \cite{luc2} and this indicates strong violation of the
Cauchy-Schwarz inequality.

The resulting expression for $\langle
\hat{A}_{j}^{\dagger}(t)\hat{A}_{j}(t)\hat{A}_{k}^{\dagger}(t)\hat{A}_{k}(t)
\rangle$ is lengthy and will not
be reproduced here.
In Fig. 5 we have plotted the quantity $I_{j,k}(t)$ indicating the
violation of Cauchy-Schwarz inequality between the $j$th mode and the $k$th
mode in dependence on  time where $(\bar{n}_{j},\lambda_{j})=(1,0.1), j=1,2,3$. Further, the solid and  short-dashed curves represent
the quantity (3.10) obtained in $(1,2)$ and $(1,3)$ modes, respectively.
For the sake of comparison the long-dashed curve is given for the single
parametric amplifier, i.e. $(\lambda_{1},\lambda_{2},\lambda_{3})=(0.1,0,0)$,
for the same values of  $\bar{n}_{j}$.
Similarly the star-centered curve is given for the single frequency converter.
In this figure the characters of both the Fock states and the structure
of the Hamiltonian, (1.4), are reflected in the behaviour of the curves.
To be more specific, the nonclassical negative values are dominant after
switching on of the interaction. Also the behaviour
of the $I_{1,2}(t)$ is smooth and becomes stable for
 large values of interaction times, however, that of the $I_{1,3}(t)$
undergoes amplification as well as oscillatory behaviour as a result of
the energy exchange between these two modes in the system (cf. (1.4)).
On the other hand, the comparison of the solid  and long-dashed curves
shows that the behaviour of the $I_{1,2}(t)$ is close to that of the single
parametric amplifier and for both  $I_{1,2}(t)\leq 0$ for all times.
This is in contrast with the behaviour of the two processes for the
sum-squeezing, see Fig. 3.
Nevertheless, this is not the case for the short-dashed (i.e.
$I_{1,3}(t)$)
and star-centered curves  where the former  can exhibit anticorrelation only
for a short time after switching on of the interaction, however, the latter
evolves periodically with period $t=2\pi/\lambda_{2}$ displaying
always anticorrelation between modes. Further,  similar behaviour can be
 seen regardless of the values of $\lambda_{j}$.
Finally, the behaviour of the  $I_{2,3}(t)$ is similar to that of
the $I_{1,3}(t)$.

\section{Quasiprobability functions}

There are three types of
quasiprobability functions: Wigner $W$-, Glauber $P$-, and Husimi $Q$%
-functions. These functions can be used  as  crucial to describe the
nonclassical effects of the system, e.g. one can employ  negative values
of $W$-function, stretching of $Q$-function and high singularities in $P$%
-function. Also, these functions are now accessible from measurements
\cite{wig}.

Indeed, the detailed statistics of the three coupled field modes can be
obtained from  photon-counting experiments. Most often we are
interested in the quantum statistics of either one mode which determine the
ensemble averages of the observables of this mode, or the composite
statistics of the compound modes which reflect their mutually correlated
properties, so that in the following we consider phase space distributions,
in particular, $W$-function
for the single- and compound-modes when all modes are initially prepared in
number states.
To achieve this goal we calculate the
$s$-parametrized joint
characteristic function  defined by

${\displaystyle C^{(3)}(\underline{\zeta },t,s)={\rm Tr}\left\{ \hat{\rho}%
(0)\exp \left[ \sum_{j=1}^{3}(\zeta _{j}\hat{A}_{j}^{\dagger }(t)-\zeta
_{j}^{*}\hat{A}_{j}(t)+\frac{s}{2}|\zeta _{j}|^{2})\right] \right\} ,}\hfill
(4.1)$

\noindent where $\hat{\rho}(0)$ is the initial density matrix operator for
the system under consideration, $\underline{\zeta}=(\zeta_{1},\zeta_{2},%
\zeta_{3})$, and $s$ takes on values $1, 0$ and $-1$ corresponding to
normally, symmetrically and antinormally ordered characteristic functions,
respectively. The superscript $3$ stands for three-mode case.
    When the modes are initially in number states $|n_{1},n_{2},
n_{3}\rangle$;  the symmetrical characteristic function can be calculated in
a straightforward way as

${\displaystyle C^{(3)}(\underline{\zeta },s=0,t)=\prod_{j=1}^{3}\exp (-%
\frac{1}{2}|\eta _{j}(t)|^{2})L_{n_{j}}(|\eta _{j}(t)|^{2}),}\hfill (4.2)$

\noindent where $L_{n}$ is the Laguerre polynomial of order $n$ and $\eta
_{j}(t)$ are given by

${\displaystyle
\eta }_{1}(t)=\zeta _{1}f_{1}(t)-\zeta _{1}^{*}f_{2}(t)-\zeta
_{2}g_{3}(t)+\zeta _{2}^{*}g_{4}(t)+\zeta _{3}h_{5}(t)-\zeta
_{3}^{*}h_{6}(t),\hfill (4.3a)$

${\displaystyle
\eta }_{2}(t)=\zeta _{2}g_{1}(t)-\zeta _{2}^{*}g_{2}(t)-\zeta
_{1}f_{3}(t)+\zeta _{1}^{*}f_{4}(t)+\zeta _{3}h_{3}(t)-\zeta
_{3}^{*}h_{4}(t),\hfill (4.3b)$

${\displaystyle
\eta }_{1}(t)=\zeta _{3}h_{1}(t)-\zeta _{3}^{*}h_{2}(t)+\zeta
_{1}^{*}f_{6}(t)-\zeta _{1}f_{5}(t)-\zeta _{2}g_{5}(t)+\zeta
_{2}^{*}g_{6}(t).\hfill (4.3c)$

We proceed by giving the definition of the $s$-parametrized joint
quasiprobability functions as

${\displaystyle
W^{(3)}(\underline{\alpha },t,s)=\frac{1}{\pi ^{6}}\int \int \int C^{(3)}(%
\underline{\zeta },t,s)\prod_{j=1}^{3}\exp (\alpha _{j}\zeta _{j}^{*}-\alpha
_{j}^{*}\zeta _{j})d^{2}\zeta _{j},}\hfill (4.4)$

\noindent where $\underline{\alpha }=(\alpha _{1},\alpha _{2},\alpha _{3})$.
When $s=1,0,-1$, the formula (4.4) gives formally Glauber $P$-function, Wigner $%
W$-function and Husimi $Q$-function, respectively.
Now substituting from (4.2) into (4.4) and performing the integration we
get the joint Wigner function as

${\displaystyle
W^{(3)}(\underline{\alpha },t)=\frac{8(-1)^{n_{1}+n_{2}+n_{3}}}{\pi ^{3}}%
\prod_{j=1}^{3}\exp (-2|\epsilon _{j}(t)|^{2})L_{n_{j}}(4|\epsilon
_{j}(t)|^{2}),}\hfill (4.5)$

\noindent where $\epsilon_{j}(t)$ are

${\displaystyle\epsilon _{1}(t)=\alpha _{1}f_{1}(t)-\alpha
_{1}^{*}f_{2}(t)-\alpha _{2}g_{3}(t)+\alpha _{2}^{*}g_{4}(t)+\alpha
_{3}h_{5}(t)-\alpha _{3}^{*}h_{6}(t),}\hfill (4.6a)$

${\displaystyle\epsilon _{2}(t)=\alpha _{1}^{*}f_{4}(t)-\alpha
_{1}f_{3}(t)+\alpha _{2}g_{1}(t)-\alpha _{2}^{*}g_{2}(t)+\alpha
_{3}h_{3}(t)-\alpha _{3}^{*}h_{4}(t),}\hfill (4.6b)$

${\displaystyle\epsilon _{3}(t)=\alpha _{1}^{*}f_{6}(t)-\alpha
_{1}f_{5}(t)+\alpha _{2}^{*}g_{6}(t)-\alpha _{2}g_{5}(t)+\alpha
_{3}h_{1}(t)-\alpha _{3}^{*}h_{2}(t).}\hfill (4.6c)$

It is  evident from (4.5)
that the joint functions are strongly affected by the correlation
between modes; this is reflected in the cross terms between different
amplitudes of different modes.
Such correlations have been used in a number of studies on nonclassical
aspects of light including questions of violation of Bell inequalities
\cite{aga}.

On the other hand, the $W$-function of the
single-mode can be obtained by means of integrating two times in the
the joint $W$-function (4.5) or by using the following relation

\begin{math}
{\displaystyle
W^{(1)}(\alpha,t)=\frac{1}{\pi^{2}}
\int
C^{(1)}(\zeta_{j},t)
\exp(\alpha\zeta_{j}^{*}-
\alpha^{*}\zeta_{j}) d^{2}\zeta_{j}, }  \hfill (4.7)
\end{math}

\noindent where  $C^{(1)}(\zeta_{j},t)$
is the single-mode symmetrical characteristic function
which can be obtained from the joint
symmetrical characteristic function (4.2).
For instance, the characteristic
function for mode $\hat{A}_{1}$ (say) can be obtained from
(4.2) by simply setting $\zeta_{2}=\zeta_{3}=0$.
Now the single-mode Wigner function for the 1st mode
when it is in the Fock states $|n_{1}\rangle$ and the rest modes
are in vacuum  can be obtained from (4.7) together with (4.2)
(after putting $\zeta_{2}=\zeta_{3}=0$  in (4.2)), carrying out
the integration, as

\begin{math}
{\displaystyle
W^{(1)}(\alpha,t)=\frac{2(-1)^{n_{1}}}{\pi\sqrt{A_{-}(t)A_{+}(t)}}
\exp\left[\frac{2h_{+}(t)}{A_{-}(t)A_{+}(t) }\right]
\sum_{l=0}^{n_{1}} \left[\frac{B_{-}(t)}{A_{-}(t)}\right]^{(n_{1}-l)}
\left[\frac{B_{+}(t)}{A_{+}(t)}\right]^{l}
}  \hfill
\end{math}

\begin{math}
{\displaystyle
\times {\rm L}^{-\frac{1}{2}}_{n_{1}-l}\left( z_{1}(t) \right)
{\rm L}^{-\frac{1}{2}}_{l}\left( z_{2}(t) \right),
}  \hfill (4.8)
\end{math}

\noindent where

${\displaystyle
A _{\pm}(t)=[f_{1}(t)\pm f_{2}(t)]^{2}+[f_{3}(t)\pm f_{4}(t)]^{2}
+[f_{5}(t)\pm f_{6}(t)]^{2},}
\hfill $

$\hfill$

${\displaystyle
B _{\pm}(t)=[f_{1}(t)\pm f_{2}(t)]^{2}-[f_{3}(t)\pm f_{4}(t)]^{2}
-[f_{5}(t)\pm f_{6}(t)]^{2},}
\hfill $

$\hfill$

${\displaystyle
C _{\pm}(t)= (\alpha^{2}+\alpha^{*2})[f_{1}(t) f_{2}(t)\pm f_{3}(t) f_{4}(t)
\pm f_{5}(t)f_{6}(t)],}
\hfill $

$\hfill$

${\displaystyle
D _{\pm}(t)= -|\alpha|^{2}[f_{1}^{2}(t)+ f_{2}^{2}(t)\pm f_{3}^{2}(t)
\pm f_{4}^{2}(t)
\pm f_{5}^{2}(t) \pm f_{6}^{2}(t)],}
\hfill $

$\hfill$

${\displaystyle
h _{\pm}(t)=C _{\pm}(t)+D _{\pm}(t),}
\hfill $

$\hfill$

${\displaystyle
z _{1}(t)=\frac{
2[A _{-}(t)+B _{-}(t)][A _{-}(t)h _{-}(t)- B _{-}(t)h _{+}(t)]
}{ A _{-}(t)B _{-}(t) [A _{+}(t)B _{-}(t)-A _{-}(t)B _{+}(t)]},
}
\hfill $

$\hfill$

${\displaystyle
z _{2}(t)=\frac{
2[A _{+}(t)+B _{+}(t)][B _{+}(t)h _{+}(t)- A _{+}(t)h _{-}(t)]
}{ A _{+}(t)B _{+}(t) [A _{+}(t)B _{-}(t)-A _{-}(t)B _{+}(t)]}
.}
\hfill (4.9)$

\noindent
 The expressions for the 2nd and 3rd modes
can be obtained from (4.8) using the transformations (2.7).

One can easily check when $t\rightarrow 0$ that the expression
(4.8) reduces to the well-known $W$-function of Fock state $|n_{1}\rangle$.
In this case the following identity \cite{[34]}

${\displaystyle\sum_{l=0}^{n_{1}
}{\rm L}_{l}^{\nu_{1}}(x){\rm L}_{n_{1}-l}^{\nu_{2}}(y)={\rm L}_{n_{1}}
^{\nu_{1}+\nu_{2}+1}(x+y)
}
\hfill (4.10)$

\noindent has to  be used.

\begin{figure}[h]%
  \centering
  \subfigure[]{\includegraphics[width=8cm]{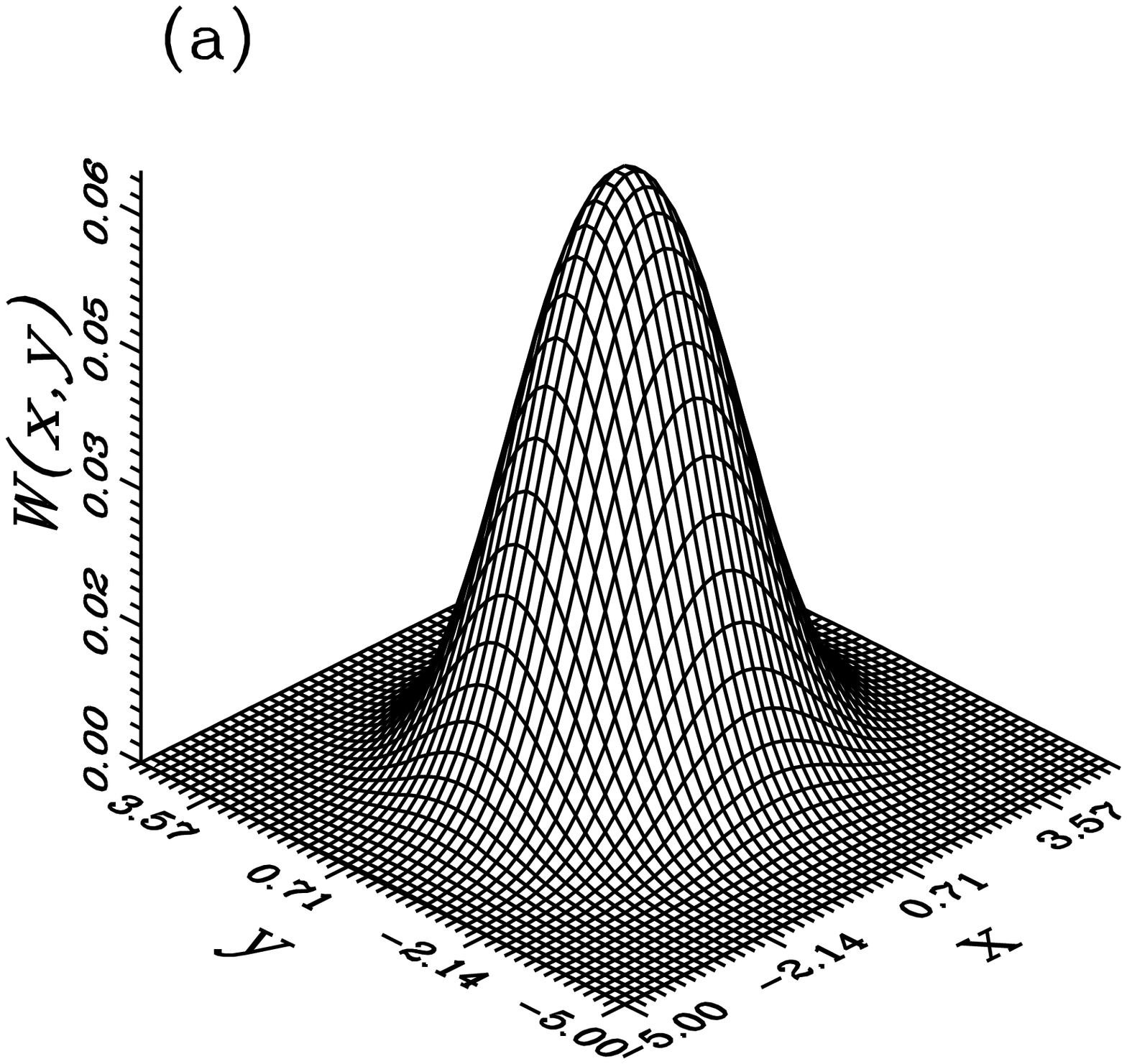}}
 \subfigure[]{\includegraphics[width=8cm]{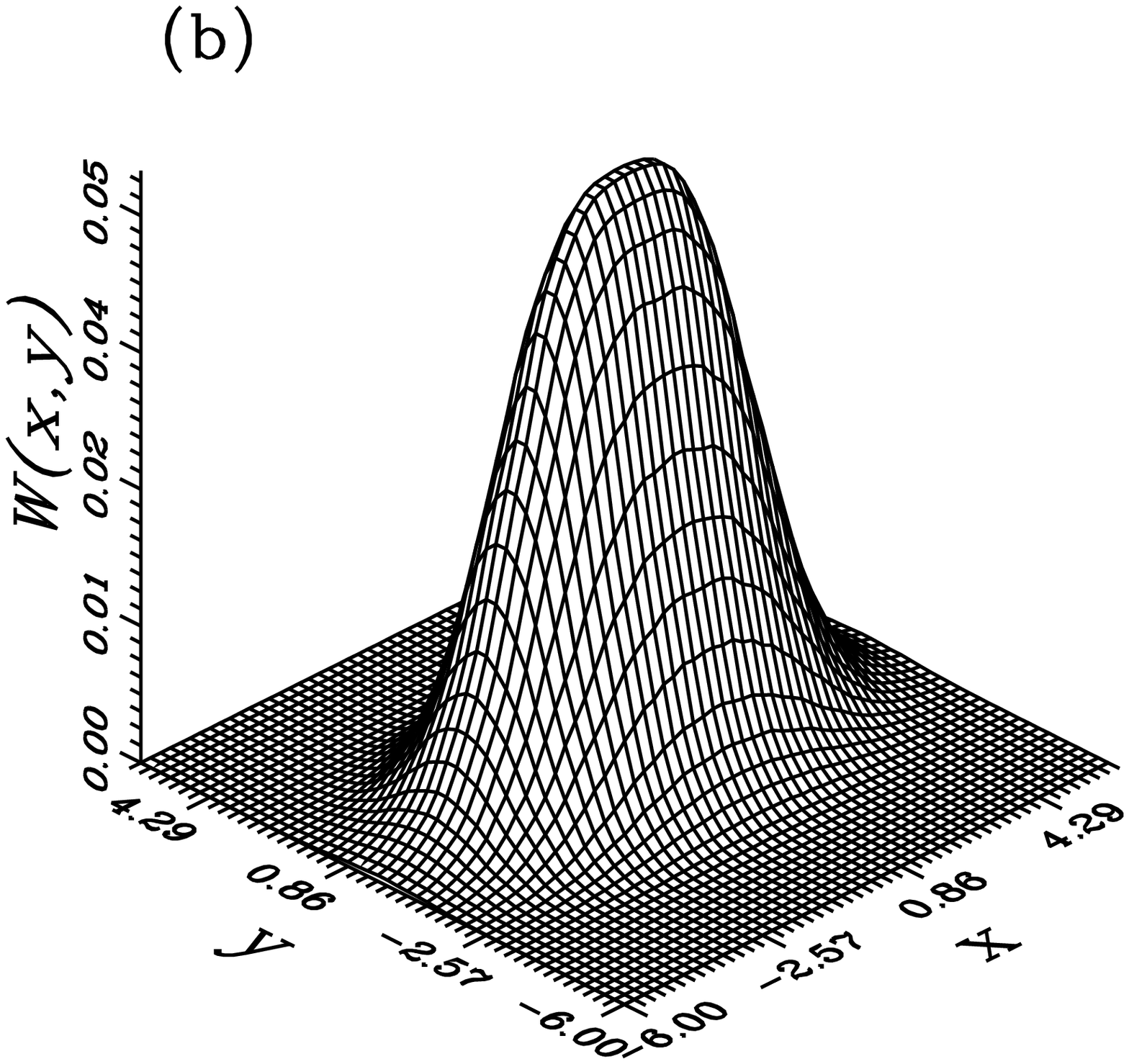}}
 \subfigure[]{\includegraphics[width=8cm]{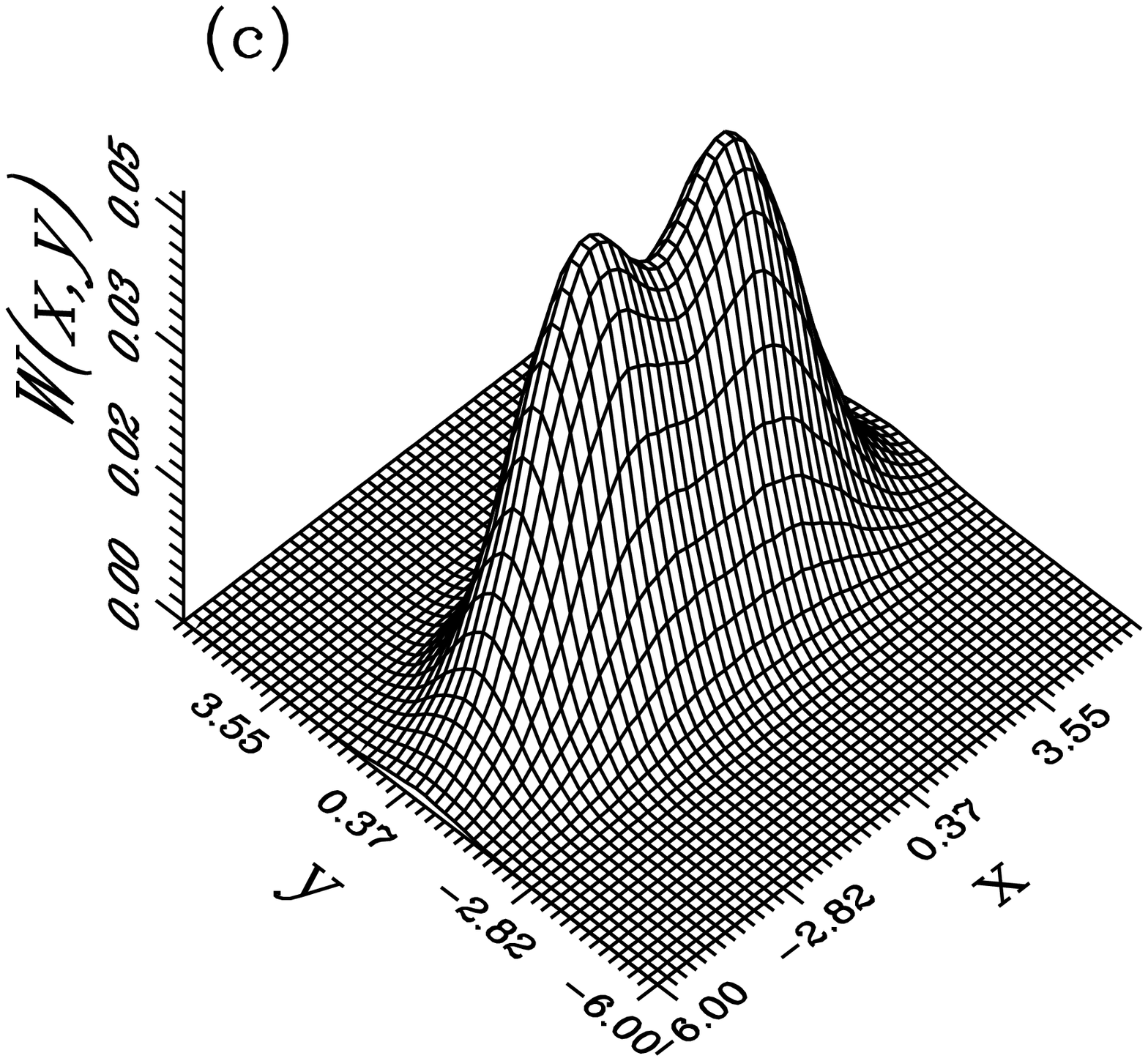}}
    \caption{
$W$-function against $x$ and $y$ for the 1st mode  when it  is in
the Fock state $|1\rangle$ and the other modes are in vacuum for
a) $(t,\lambda_{1},\lambda_{2},\lambda_{3})
=(\frac{\pi}{2},1.005,1,0.1)$ ; b)
$(t,\lambda_{1},\lambda_{2},\lambda_{3})
=(\frac{\pi}{2},1.008,1,0.4)$ ; c)
$(t,\lambda_{1},\lambda_{2},\lambda_{3}) =(\pi,0.5,0.4,0.4)$.  }
  \label{fig6}
\end{figure}

Now we proceed by studying the evolution of the $W$-function for the 1st mode when it is
initially in the Fock state $|1\rangle$ and the other modes are in vacuum
using the expression (4.8). In
fact, the $W$-function of the Fock state $|1\rangle$ is well known and it
exhibits an inverted hole with more pronounced negative values related to
the state showing maximum sub-Poissonian statistics. However, this behaviour
can be completely washed out in our model at certain values of interaction
parameters. This can be seen in Fig. 6a where one can observe the
well-known shape for the $W$-function of the thermal light, which is broader
than that for vacuum and this demonstrates the Bose-Einstein statistics for
thermal light exhibiting larger fluctuations than for coherent light. So the
competition between different states in the model (power transfers) can
cause that the output field exhibits classical nature of thermal light. In
Fig. 6b, by increasing the values of the coupling constants, $W$-function
for squeezed thermal light dominates. This result is different from that
given in
section 3 since there the initial thermal fields undergo squeezing when
evolve in the present structure. It is important to mention that stretching
here can be more pronounced for large interaction time. Finally, Fig. 6c
displays an interesting fact that a statistical mixture for coherent
superposition
states, in principle, can also be realized where the basis of the two
Gaussian bells dominates in the behaviour of $W$-function. This behaviour
may be achieved also from the cat states when they interact with a finite
temperature heat bath \cite{buz4} or with the bath which consists of a gain
medium in addition to the usual absorber \cite{agar}. For example, in the
latter case this can be realized by choosing the gain appropriately so that
one gets a purely diffusive motion of the field mode which leads to a double
Gaussian structure with the missing oscillatory behaviour. Thus based on the
fact that the nonlinear coupling between modes is present in the medium, the
fields which are initially nonclassical cannot recover their properties
during the evolution in the model for strong coupling and large interaction
time. Nevertheless, they lose some nonclassical features, e.g. the $W$%
-function loses its negative values, and is obtaining other ones, e.g.
stretching of its form.

\begin{figure}[h]%
    \includegraphics[width=8cm]{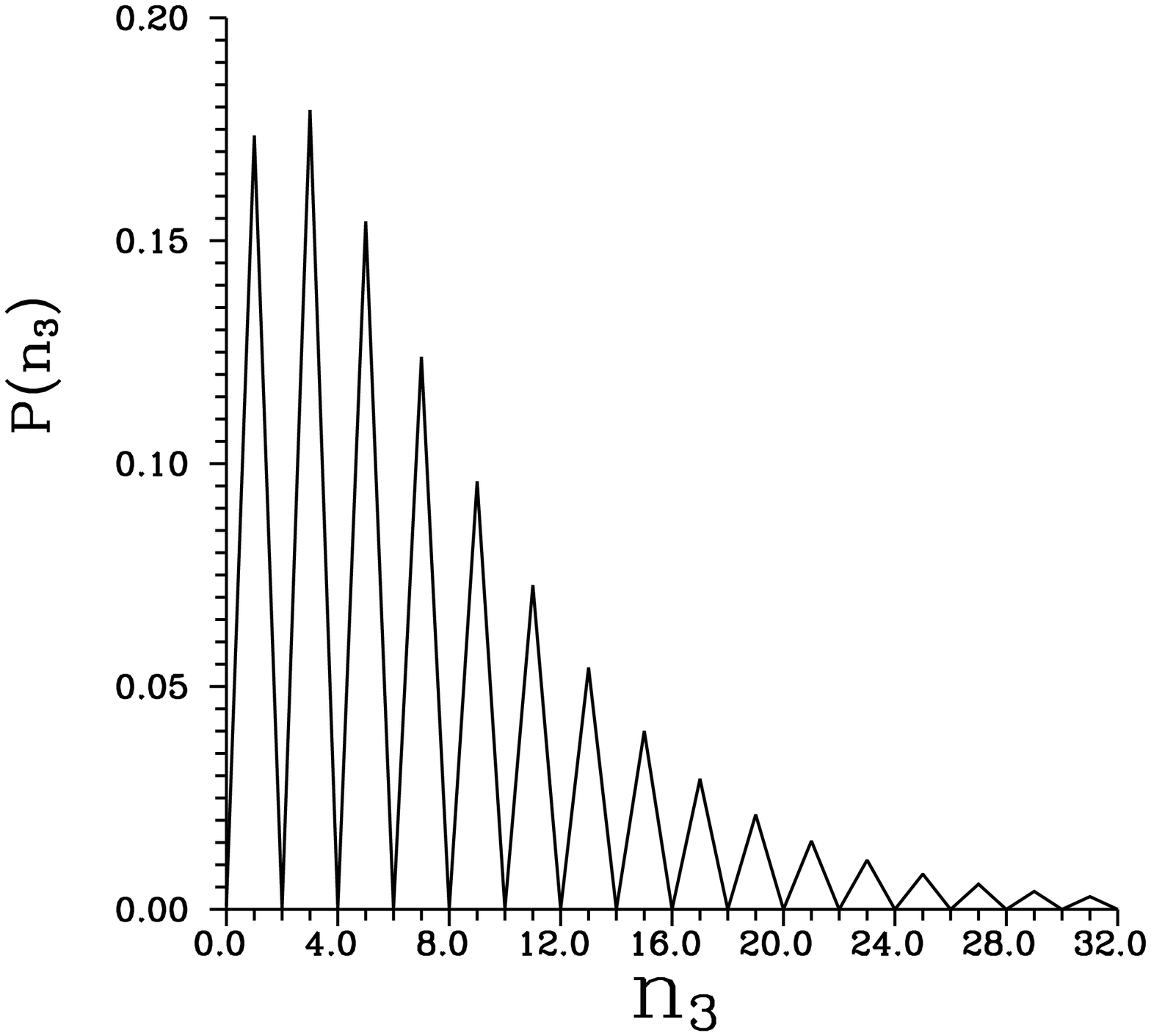}
   \caption{
hoton-number distribution $P(n_{3})$ for the 3rd mode   when it is
initially  in the Fock state $|1\rangle$ and the other modes are
in vacuum for $t=8$ and $(\lambda_{1},\lambda_{2},\lambda_{3})
=(0.3,0.3,0.3)$.
  }
  \label{fig7}
\end{figure}


Finally, we would like to close this section by paying attention to the
behaviour of the photon-number distribution when the modes are initially
in Fock states.
This will be done by means of  the
integral expression for the photon-number distribution
 in terms of Wigner function and Laguerre polynomials,
 which is given by

${\displaystyle
P(n_{j})=\frac{2(-1)^{n_{j}}}{n_{j}!}\int W^{(1)}(\alpha)\exp
(-2|\alpha|^{2})L_{n_{j}}(4|\alpha |^{2})d^{2}\alpha ,}\hfill
(4.11)$

\noindent where $j$ denotes the mode under consideration and $
W^{(1)}(\alpha)$ is  the single-mode Wigner function.
We restrict ourselves to the photon-number distribution  for the 3rd mode
when it is initially in Fock state $|1\rangle$ and the other modes are
in vacuum.    After inserting the expression for $W$-function of this
case into (4.11) and performing the integral,  the
 result has been plotted in Fig. 7 for shown values.
In this figure one can see the oscillatory behaviour typical for
the photon-number distribution of squeezed  states in the 3rd mode.
This is clear in the  pairwise  nature of the oscillations, i.e. $P(2n_{3})=0,
 \quad n_{3}$ is positive integer. Indeed, for
the  squeezed states, i.e. squeezed vacuum, squeezed number and
squeezed thermal states, these pairwise oscillations are explained
as the result of the quadratic,
or two-photon nature of the squeeze operator $\hat{S}(r)$ \cite{yun2}.

\section{Conclusion}

In this article we have studied the statistical properties of three harmonic
oscillators mutually interacting in the nonlinear crystal. The model is
governed by interaction Hamiltonian including competition between
 two frequency converters and one parametric amplifier.
After using the Heisenberg approach to the quantum statistics
of interacting modes, we have investigated effects produced by the dynamics
of the interaction as well as by the nonclassical behaviour of the initial
light modes. In our analysis we have considered that the modes are
initially in  the number states. We have proved  that squeezed light can
be generated in the standard sense or in the correlated quadratures (sum-squeezing).
Furthermore, we have
shown that this is still valid if the Fock states are replaced by
thermalized  fields. In this case the model
operates as a microwave Josephson-junction parametric amplifier. We have
also demonstrated that the initial sub-Poissonian statistics of the
interacting modes are washed out, which leads to partially coherent, chaotic
and superchaotic fields. Finally, we have shown that when the system is
initially prepared in the Fock state $|1,0,0\rangle$, at certain values of
interaction parameters, the 1st mode can evolve into thermal states or
in squeezed thermal states.

{\bf Aknowledgments}

We thank Prof. V. Pe\v{r}inov\'{a} and Dr. A. Luk\v{s} from Department of
Optics, Palack\'y University, Olomouc, Czech Republic for the critical
reading of the article.
J. P. and F. A. A. E-O. aknowledge the partial support from the Projects
 LN00A015 and Research Project CEZ: J 14/98 of Czech Ministry of Education
and 202/00/0142 of Czech grant agency.
One of us (M. S. A.) is grateful for the financial support
from the Project Math 1418/19 of the Research Centre, College of
Science, King Saud University.

\end{document}